\documentstyle[12pt]{article}

\makeatletter                                                           

\def\section{\@startsection {section}{1}{\z@}{-1.5ex plus -.5ex         
minus -.2ex}{1ex plus .2ex}{\large\bf}}                                 

\def\@thmcountersep{}                                                   

\makeatother

\begin{document}

\begin{titlepage}
\vspace{1cm}
\begin{flushright}
gr-qc/9409021\\
\end{flushright}
\vspace{1.5cm}
\begin{center}
{\LARGE \bf What Cavities?} \\
\vspace{1.2cm}
{\large Joseph Katz\footnote{Permanent address: Racah Institute of Physics,
    Hebrew University, 91904 Jerusalem, Israel}\footnote{e-mail:
    jkatz@hujivms.bitnet} 
and Yasunari Okuta\footnote{e-mail: okuta@gprx.miz.nao.ac.jp}
\vspace{0.4cm}} \\
{\normalsize \it Division of Theoretical Astrophysics, National  \\
\small \it Astronomical Observatory, Mizusawa, Iwate 023, Japan}

\vspace{2cm}
{\large \bf Abstract}
\end{center}
 Most spherical thin shells, enclosing black body radiation satisfy
the dominant energy condition if they have at least $\simeq 30\%$ of the total
mass-energy.
 Containers with less mass energy, able to sustain high pressures, contain
mostly unstable radiation. If they have negligible mass energy they are
unable to sustain the pressures and the radiation is unstable to gravitational
collapse.
 Containers with black holes and radiation in thermal equilibrium,
considered in the literature, are often unrealistic.
\end{titlepage}

\section{Introduction}

 Thermal equilibrium of black holes and radiation may exist in closed
cavities (for recent reviews on the subject see papers by Landsberg
\cite{Land:92} and Page \cite{Page:92}).
The walls must be perfectly reflecting and able to resist pressures or tensions
upon demand.
Or so it seems for the resistance of the box is rarely questioned.
The mass of the wall is ignored or explicitly neglected like in
Sorkin, Wald and Zhang \cite{SWZ:81} though far reaching conclusions are
sometimes drawn.
\par
 From an engineering view point, cavities with properties as asked for in
black hole thermodynamics are incredible.
Forces in the cavity must equilibrate the pressure due to radiation, attenuated
by self-gravity attraction.
But radiation pressure is of the order of its mass energy density.
 Therefore, how can a cavity have negligible mass energy (compared to that of
radiation) and at the same time resist pressure forces comparable to
those of radiation (divided by the radius of the box $R$)?
 Surely some \lq energy condition\rq \ is strongly violated.
\par
 To shed some light on this question we have calculated the mass $M_s$ of
cavities with thin walls (thin spherical shells) and surfaces equal to
$4 \pi R^2$.
 The force $F$ on half a sphere satisfies the dominant energy condition
\cite{HE:73} if
\begin{equation}
     - M_s c^2/ 2 R \leq F \leq M_s c^2/ 2 R.
\label{eq:dec}
\end{equation}
 The dominant energy condition is the only one to restrain pressures
as well as tensions.
 One should not allow for unlimited pressure, this is not physical.

 The results of this calculation are summarized here in broad terms.
 Finer details are given in the text.
 Let $\eta$ be the fraction of total mass-energy in the cavity ($\eta<1$).
 When $\eta<30 \% $ there are two different types of equilibrium configurations
which are both unviable but for different reasons.
 There are cavities very close to the Schwarzschild radius with pressures so
high that they do not satisfy eq.(\ref{eq:dec}).
 And there are cavities with bigger radii which satisfy eq.(\ref{eq:dec}) but
in which the radiation is either unstable and on the verge of gravitational
collapse.
 In practice, realistic cavities with stable radiation must have more than
$30\%$ of the total mass energy.

 For cavities with $\eta>30\%$ there is another novelty.
 The inverse temperature $\beta $ as a function of $R$ has no succession
of maxima and minima anymore as is typical of equilibrium configurations
with equations of state of the form
$P \propto \rho$ \cite{Chandra:72}.
 $\beta$ has only one minimum.
 The energy of the radiation has only one maximum beyond which
no equilibrium configuration exists.
 That maximum is also associated with a limit of
thermodynamic stability.
 The stability limit is smaller for higher $\eta$'s.
 Thus, radiation collapses into a black hole \cite{GP:78} at smaller energies
than thought previously \cite{PKO:94}.
 These are the main features of \lq viable \rq \ cavities, those that are
stable and do not violate the dominant energy condition.

\vspace{0.6cm}
\section{Elements of the Problem}

 We consider spherical cavities with perfectly reflecting walls enclosing
black-body radiation (photons).
 The spacetime outside is thus Schwarzschild and the metric can be written in
terms of the total mass energy $M$ or $m=GM$ (with $c=1$) as
\begin{eqnarray}
 ds^2=\left(1-\frac{2m}{r} \right) \: dt^2
     - \left(1-\frac{2m}{r} \right)^{-1} \: dr^2
     - r^2 d\Omega^2,& & \makebox[0.7cm]{} r \geq R \\
    d\Omega^2\equiv d\theta^2+\sin^2\theta \, d\phi^2\; & &\nonumber
\label{eq:metricout}
\end{eqnarray}
 $R$ is the \lq radius\rq \ of the cavity.
 The radiation inside has pressure $P$, mass energy density $\rho$ and local
temperature $T$ which all vary with $r$ and, in Planck units, are related
as follows:
\begin{equation}
 P=\frac13 \rho, \;\;\; \rho=\frac{\pi^2}{15}\:T^4
\label{eq:pr}
\end{equation}
 The solution of Einstein's equations is well known, having been studied by
Klein \cite{Klein:47}, Chandrasekhar \cite{Chandra:72}, Sorkin, et.al.
\cite{SWZ:81}, Landsberg \cite{Land:92} and Page \cite{Page:92}.
 Solutions are scale free and all quantities can be expressed in units of R.

 The metric outside can be written as
\begin{equation}
 ds^2=e^{2\nu}dt^2 - \frac{dr^2}{1-2 \mu(r)} - r^2 d\Omega^2,
    \makebox[1.2cm]{} r\leq R
\label{eq:metricin}
\end{equation}
 With all clocks referred to the proper time at infinity $t$, we have a
junction
condition on $\nu(R)$:
\begin{equation}
e^{2\nu(R)}=1-\frac{2m}{R},
\label{eq:jc1}
\end{equation}
elsewhere, $\nu(r)$ is given by
\begin{equation}
 e^{4\nu(r)}\, \rho(r)=e^{4\nu(R)}\, \rho(R)
\label{eq:jc2}
\end{equation}
where
\begin{equation}
\rho(R)=\frac{\pi^2}{15}\: T^4(R)  \;\;\;\;\; \mbox{and} \;\;\;\;\;
  \sqrt{1-\frac{2m}{R}}\: T(R)=T_\infty =\frac 1 \beta
\label{eq:jc3}
\end{equation}
Given $\beta$ and $R$, (\ref{eq:jc3}) fixes the values for $T(R)$ and
$\rho(R)$.
$\rho(r)$ determines not only $\nu(r)$ but also $\mu(r)$:
\begin{equation}
\mu(r)=\frac{\bar m(r)}{r}, \;\;\;\;\; \bar{m}(r)=\int^r_0 4 \pi r'^2 \rho'
dr'.
\label{eq:defmu}
\end{equation}
$\rho(r)$ itself is obtained from Einstein's equations.
Setting like in Sorkin et al. \cite{SWZ:81}
\begin{equation}
q(r)=\frac{d\bar{m}}{dr}=4 \pi r^2 \rho \;\;\;
  \mbox{and} \;\;\; z=\ln r,
\label{eq:defq}
\end{equation}
one has
\begin{eqnarray}
 \frac{d\mu}{dz}&=&q-\mu, \\
\label{eq:dmudz}
 \frac{dq}{dz}&=&\frac{2q[1-\frac 23 q-4\mu]}{1-2\mu}.
\label{eq:dqdz}
\end{eqnarray}
 From these two equations one can eliminate z and obtain the
Tolman-Oppen-heimer-Volkoff equation for $q(\mu)$:
\begin{equation}
\frac{dq}{d\mu}=\frac{2q\left(1-\frac 23 q-4\mu\right)}{(1-2\mu)(q-\mu)}
   \;\;\;\;\; \mbox{with} \;\;\;\;\; q=\mu=0
\label{eq:tov}
\end{equation}
 This equation is easily integrated; $q(\mu)$ has been given in other works,
but
it is useful to have $q(\mu)$ reproduced here (figure 1).
 With $q(\mu)$ one obtain $q(r)$ by integrating (\ref{eq:dqdz})
\begin{equation}
\frac Rr = \exp
   \left[\int^{q_B}_q \,\,\frac{(1-2\mu)\: dq}{2q(1-\frac 23 q-4\mu)} \right],
\label{eq:Rr}
\end{equation}
in which
\begin{equation}
q_B=4 \pi R^2 \rho_B
\label{eq:qB}
\end{equation}
and $\rho_B$ is defined by $R$ and $\beta$ as we have seen.
 Finally if we have $q(r)$ we obtain $\rho(r)$ from eq.(\ref{eq:defq}).
$\nu(r)$ is thus completely defined and depends on $\beta$ and on
\begin{equation}
\alpha=\frac mR.
\label{eq:delalpha}
\end{equation}
$\nu(r)$ is independent of the shell, but $\mu(R)=\bar {m}(R) / R$
depends on the mass energy of the cavity.

\vspace{0.6cm}
\section{The Cavities}

 Let $\sigma$ be the mass energy density of the thin sphere enclosing the
radiation and $\Pi$ the pressure ($\Pi > 0$) or tension ($\Pi < 0$) in the
shell.
 The relation between $\sigma$, $\Pi$ and the metrics on both sides are
readily found from second fundamental forms and from Gauss-Codazzi identities
\cite{Israel:66}.
 Following for instance Goldwirth and Katz \cite{GK:94}, one finds that
\begin{equation}
 4 \pi G R \sigma = \sqrt{1-2\mu} - \sqrt{1-2 \alpha}, \;\;\;\;\; \mu=\mu(R)
\label{eq:defsig}
\end{equation}
and
\begin{equation}
 4 \pi G R \Pi = \frac 12
   \left [ \frac{1-\alpha}{\sqrt{1-2 \alpha}}
          -\frac{1-\mu+\frac 13 q(\mu)}{\sqrt{1-2 \mu}} \right].
\label{eq:defpi}
\end{equation}
 Other useful expressions may be derived in terms of global quantities.
 First, in terms of the mass $M_s=4 \pi R^2 \sigma$ of the shell
for (\ref{eq:defsig})
\begin{equation}
 \alpha_s = \frac{GM_s}{R} = \frac{m_s}{R}
          = \sqrt{1-2\mu} - \sqrt{1-2\alpha}.
\label{eq:defas}
\end{equation}
 Second, in terms of the force $F$ at the rim of half a sphere:
\begin{equation}
 F = 2 \pi R \: \Pi,
\label{eq:defF}
\end{equation}
equation (\ref{eq:defpi}) becomes
\begin{equation}
 GF = f = \frac 14 \left [ \frac{1-\alpha}{\sqrt{1-2 \alpha}}
                          -\frac{1-\mu+\frac 13 q}{\sqrt{1-2\mu}} \right].
\label{eq:defGF}
\end{equation}
The dominant energy condition $\sigma>0$ and $-\sigma<\Pi<\sigma$ amounts thus
to $\alpha_s>0$ and
\begin{equation}
    -\frac{\alpha_s}{2}<f<\frac{\alpha_s}{2},
\label{eq:dec2}
\end{equation}
or in more physical term, to eq.(\ref{eq:dec}) if $c$ is not set equal to $1$.
 One straightforward consequence of equation (\ref{eq:defas}) and $\alpha_s>0$
is that
\begin{equation}
 \mu(R) < \alpha.
\label{eq:muLEa}
\end{equation}

\vspace{0.6cm}
\section{Equilibrium Configurations and Mass Energy of the Shells}

 $\alpha_s$ does not represent the mass energy of the shell $E_s$ which is
given by
\begin{equation}
 E_s = \int \! \sqrt{-g} \; {T^0_0}_{shell} \: d^3x
     = \sqrt{1-2\alpha} \, M_s.
\label{eq:defEs}
\end{equation}
 Thus, in units of $R$,
\begin{equation}
 \frac{E_s}{R} = \alpha_s \sqrt{1-2\alpha}.
\label{eq:EsovR}
\end{equation}
 It follows from (\ref{eq:defas}) that
\begin{equation}
 \frac{E_s}{R} =
    \sqrt{1-2\alpha} \left( \sqrt{1-2\mu} - \sqrt{1-2\alpha} \right)
        \leq \sqrt{1-2\alpha} \left(1-\sqrt{1-2\alpha} \right) \leq \alpha
\label{eq:EsLEa}
\end{equation}
and therefore $E_s\leq M$. The fraction of mass energy in the shell is defined
by
\begin{equation}
    \eta = \frac{E_s}{M} \leq 1.
\label{eq:etaLE1}
\end{equation}
 For a given $\eta$, equation (\ref{eq:EsovR}) relates $\mu$
to both $\alpha$ and $\eta$.
 There is some advantage in introducing new variables instead of $\mu$ and
$\alpha$:
\begin{equation}
    y \equiv \sqrt{1-2\mu}, \;\;\; z \equiv \sqrt{1-2\alpha},
\label{eq:defyz}
\end{equation}
because of eq.(\ref{eq:muLEa}) $0 \leq z \leq y \leq 1$.
 Thus eq.(\ref{eq:EsovR}) for $\mu(\alpha,\eta)$ is now replaced by an equation
for $y(z, \eta)$:
\begin{equation}
 y = \frac{\eta/2}{z} + \left( 1-\frac{\eta}{2} \right) z.
\label{eq:relyz}
\end{equation}

 We know from figure 1 that equilibrium configurations exist only for
\begin{equation}
 \mu \leq \mu_{max}=0.246 \simeq \frac 14 \;\;\; \mbox{or}
      \;\;\; y(\mu_{max}) \geq \frac{1}{\sqrt{2}}.
\label{eq:limmuy}
\end{equation}
 This limit $1/\sqrt{2}$ divides the equilibrium configurations into two
classes depending on whether the parabolic curve $y(z)$ in eq.(\ref{eq:relyz})
has a minimum below or above $1/\sqrt{2}$.
The minimum of $y$ is
\begin{equation}
 y_{min} = \sqrt{\eta \, (2-\eta)}.
\label{eq:ymin}
\end{equation}
 The dividing line at $\eta=\eta_0$ corresponds thus to
$\sqrt{\eta_0\,(2-\eta_0)}=\frac{1}{\sqrt{2}}$ or
\begin{equation}
 \eta_0 = 1-\frac{1}{\sqrt{2}} \simeq 0.293 \simeq 0.3.
\label{eq:eta0}
\end{equation}
 If $\eta < \eta_0$, then $y_{min} < y(\eta_0)$ (figure 2).
 Equilibrium configurations exist for all values of $\mu$:
$0\leq \mu\leq \mu_{max}$ but in two separate ranges of $z$.

 The first range
\begin{equation}
 z_1 = \frac{\eta}{2-\eta} \leq z
     \leq \frac{1-\sqrt{1-2\eta(2-\eta)}}{\sqrt{2}\: (2-\eta)} = z_2
\label{eq:rangez1}
\end{equation}
is for small values of $z$.
 Indeed both $z_1$ and $z_2$ are monotonic functions of $\eta$ and the interval
is
always close to the Schwarzschild radius, ranging from $(0,0)$ for $\eta=0$ to
$(z_1=0.172, z_2=0.414)$ for $\eta=\eta_0$, which corresponds to
$\frac{R_1}{2m}=1.03 \leq \frac{R}{2m} \leq 1.21=\frac{R_2}{2m}$.

 The second range of $z$ is
\begin{equation}
 z_3 = \frac{1+\sqrt{1-2\eta(2-\eta)}}{\sqrt{2}\: (2-\eta)} \leq z \leq 1
\label{eq:rangez3}
\end{equation}
and has a big range with $\frac{R_3}{2m} \leq \frac{R}{2m} \leq \infty$,
the lower limit being between $\frac{R_3}{2m}=1.21$ for $\eta=\eta_0$ and
$\frac{R_3}{2m}=2$ for $\eta=0$.

 If $\eta > \eta_0$, $y_{min}$ is greater than $y(\mu_{max})$ and equilibrium
configurations do not exist for $\mu$ greater than some
$\tilde{\mu}(\eta) < \mu_{min} \simeq \frac 14$.
 However, the range for $z$ is now continuous (see again figure 2):
\begin{equation}
 z_1 = \frac{\eta}{2-\eta} \leq z \leq 1.
\label{eq:z1LE1}
\end{equation}
 With $\eta>\eta_0$, $z_1\geq 0.172$ and (\ref{eq:z1LE1}) corresponds to
a total range $1.03 \leq \frac{R}{2m} \leq \infty$, almost down to
the Schwarzschild limit. \par

\vspace{0.6cm}
\section{Linear Series for Radiative Equilibrium}

 The three classes of solutions correspond to quite different $\beta(\mu)$
lines of equilibrium configurations.
 These lines can be deduces from eq.(\ref{eq:jc3}) and (\ref{eq:defq}) which
give
\begin{equation}
  B \equiv \left(\frac{15}{4\pi^3} \right) \frac{\beta}{R^{1/2}}
    = \frac{1}{zq^{1/4}}
\label{eq:defB}
\end{equation}
with $q(\mu)$ and $z(y,\eta)$ defined by eq.(\ref{eq:relyz}) and with
$y=\sqrt{1-2\mu}$, (\ref{eq:defB}) gives $\beta(\mu,\eta)$ at fixed $R$.

 When $\eta<\eta_0$, consider first the class of equilibrium configurations for
which $z$ varies in the interval $(z_1,z_2)$.
 Here $\mu$ grows from $0$ at $z=z_1$ to $\mu_{max}$ at $z=z_2$
and $q$ varies accordingly, as can be seen on figure 1, from point O to P.
 But figure 1 shows also that equilibrium configurations exists for
decreasing $z$'s, decreasing $\mu$'s and {\it further decreasing} $q(\mu)$.
 Thus for $\eta<\eta_0$ and $z_1 \leq z \leq z_2$, $z$ and $y$ may oscillate
near $z_2$ a number of times for which $B(\mu)$ varies just as shown in
figure 3a, a result of $q(\mu)$'s behavior.
 The class of equilibrium for $z_3 \leq z \leq 1$ are similar to those of
figure 3a but in a different range of temperature as can be seen in
figure 3b where $\eta=0$ is also drawn.
 The limit $\eta\rightarrow 0$, taken by Sorkin et al., amounts to neglect
the cavity.
 Both curves of figure 3 are inward spiralling characteristic
of equations of state of the form $P/\rho=$const. \cite{Chandra:72}.

 When $\eta>\eta_0$, equilibrium configurations exist only for
$0 \leq \mu \leq \tilde{\mu}(\eta) \leq \mu_{max}$.
 However, when $\mu$ varies from $0$ to $\tilde{\mu}$, $z$ varies from $z_1$
to $\sqrt{z_1}$ where $y=y_{min}$ and with increasing value of $z$, $\mu$
decreases again from $\tilde\mu(\eta)$ to zero.

 The corresponding linear series $B(\mu)$ is an unwound spiral shown in
figure 4.
 Here $B(\mu)$ tends to infinity twice, for $z=z_1$ and $z=1$, since in both
limits $y\rightarrow 1$, $\mu\rightarrow 0$ and $q \rightarrow \frac 13 \mu$
so that $q\rightarrow 0$ and $B\rightarrow \infty$.

\vspace{0.6cm}
\section{Viable Cavities}

 Among all equilibrium configurations of given $\eta$, what are those that
do not violate conditions (\ref{eq:dec2})?
 Or equivalently, in what region of the $(y,z)$ plane or the $(\mu(R),\alpha)$
plane or the $(\bar{m}(R),m)$ plane does (\ref{eq:dec2}) hold for
$0 \leq \eta \leq 1$?
 If in equation (\ref{eq:dec2}) we replace $f$ as given by eq.(\ref{eq:defGF})
and $\alpha_s$ by eq.(\ref{eq:defas}) and use $(y,z)$ coordinates defined in
(\ref{eq:defyz}) we obtain the following inequalities:
\begin{equation}
  -\frac 12 (y-z) \leq f=\frac 18 (\frac{1}{yz}-1)(y-z)-\frac{q}{12y}
    \leq \frac 12 (y-z).
\label{eq:ineqfyz}
\end{equation}
 These inequalities tell us also that
\begin{equation}
   \left(\frac 1z -5y\right)(y-z) \leq \frac{2q}{3}
     \leq \left(\frac 1z +3y\right)(y-z).
\label{eq:ineqqyz}
\end{equation}
 The left hand side is a limit on pressures ($\Pi<\sigma$), the right hand side
on tensions ($\Pi>-\sigma$). \par

\subsection{\rm Limits on the pressures}
 Normal cavities must resist the radiation pressure. Pressure turns into
tensions when self-attraction of the shell and radiation becomes important.
 Therefore, ordinary cavities are submitted to {\it tension} while pressure
in the wall of the cavity only appear in extremely relativistic case.

 The left hand inequality (\ref{eq:ineqqyz}) corresponding to $\Pi<\sigma$
can be written with $y$ replaced by (\ref{eq:relyz}) and reads then
\begin{equation}
  \left[\left(\frac 25-\eta\right) - (2-\eta) z^2 \right]
        \left(1-z^2\right)
    < \frac{8q}{15\eta} z^2.
\label{eq:ineqqzeta}
\end{equation}
 Since $q \geq 0$, (\ref{eq:ineqqzeta}) always holds if the left hand side is
negative, that is,
\begin{equation}
   z^2 > \frac{\frac 25 -\eta}{2-\eta}
\label{eq:ineqzeta}
\end{equation}
(\ref{eq:ineqzeta}) is certainly satisfied for $\eta > \frac 25 = 0.4$, but
for a given $\eta$, $z \geq z_1 = \eta / (2-\eta)$.
 As a result, (\ref{eq:ineqzeta}) will hold for all $z$ if it holds for $z_1$,
that is, when $\eta > \frac 13$ which is of the same size as
$\eta_0 \simeq 0.3$.
 Thus for $\eta_0 > \frac 13$ the pressure in the wall of the cavity is never
too high.

 Another piece of information follows from (\ref{eq:ineqzeta}) for
$y \rightarrow 1$ , that is, $\mu \rightarrow 0$, and $z \rightarrow 1$ or to
$z_1$ (see figure 2).
 Since for $\mu \ll 1$, $q \simeq 3\mu$, one also has
\begin{equation}
 q \simeq 3\mu = \frac 32 (1-y^2) \simeq 3 (1-y),
\label{eq:simqmuy}
\end{equation}
because $1+y \simeq 2$.
 We now replace $q$ by (\ref{eq:simqmuy}) into (\ref{eq:ineqqzeta}):

\begin{enumerate}
\item For $z \rightarrow 1$, $M \rightarrow 0$, (\ref{eq:ineqqzeta}) becomes
\begin{equation}
  -2 \eta(1-z) < (1-y).
\label{eq:ineqetazy}
\end{equation}
 This is always satisfied.
 The pressure is thus never too high in big cavities ($R\gg M$).

\item For $z \rightarrow z_1=\eta/(2-\eta)$, $R/2m$ is small and
(\ref{eq:ineqqzeta}) becomes
\begin{equation}
  \left(\frac 13 -\eta\right)(1-\eta) < \frac 16 \eta (2-\eta)(1-y)
\label{eq:ineqetay}
\end{equation}
which is never satisfied when $\eta < \frac 13$ and $y \rightarrow 1$.
 The pressure sustained by the cavity becomes thus always too high for
$R\rightarrow R_1$ when $\eta<1/3$.
 (\ref{eq:ineqetay}) is always correct for $\eta > \frac 13$, but this we
already know.
\end{enumerate}

\subsection{\rm Limits on tensions}
 Limits on tensions are obtained from the right hand side of
(\ref{eq:ineqqyz}).
 With $y$ replaced by (\ref{eq:relyz}) one has
\begin{equation}
 \frac{8q}{9\eta} < \frac{1}{z^2} \left[\left( \frac 23 + \eta \right)
   + \left(2-\eta\right) z^2 \right] (1-z^2).
\label{eq:ineq89}
\end{equation}
 When $y\rightarrow 1$ ($\mu \rightarrow 0$) and $q \simeq 3(1-y)$
(see (\ref{eq:simqmuy})), then

\begin{enumerate}
 \item For $z\rightarrow 1$, $M\rightarrow 0$, (\ref{eq:ineq89}) holds only
if $\eta>\frac 13$.
 Thus when $\eta>\frac 13 \simeq \eta_0$, tensions are never too high.
 However, for $\eta<\frac 13$ and thus also for shells with negligible mass
energy, tensions are always too high in big enough cavities.
 This means that even when gravity is weak there is always a non negligible
lower limit to the mass energy of the cavity.

 \item For $z \rightarrow z_1$, (\ref{eq:ineq89}) becomes
\begin{equation}
    (z-z_1) < \frac{2(1+\eta)}{(2-\eta)^2}.
\label{eq:z-z1}
\end{equation}
Since $z-z_1 \rightarrow 0$, this inequality is always satisfied and tensions
are never too big as $R\rightarrow R_1$.
\end{enumerate}

 The limits on tensions or pressures correspond thus approximately to the
dividing line between the 2 types of solutions ($\eta<\eta_0$ or
$\eta>\eta_0$).
 The class of cavities with $30\%$ or more mass energy in the shell can resist
any pressure or tension in all admissible equilibrium configurations.
 On the other hand tensions are always too high in light cavities ($\eta<30\%$)
with negligible self-gravity ($z\rightarrow 1$).
 Cavities with negligible mass-energy explode.
 And pressures are always too high in light cavities close to the Schwarzschild
limit.
 Such cavities collapse.
 Figure 5 shows the region in the $(\alpha,\mu)$ plane in which $|\Pi|<\sigma$
(see also figure 3).
 It also shows the limited domain of existence of the 2 classes of equilibrium
for $\eta<\eta_0$.
 It is clear that most of these equilibrium configurations have no viable
cavities.

\vspace{0.6cm}
\section{Stability Limits}

 Thermodynamic stability limits for spherical perturbations are reached when
the radiation energy $E_r$ has a turning point in a linear series of
equilibrium configurations.
 If the mass energy and field energy outside of the cavity are negligible,
$E_r/R\simeq\mu(R)\simeq \alpha$, configurations are stable for
$0<\mu<\mu_{max}=0.246$ (see figure 3b)\cite{PKO:94}.
 $\mu_{max}$ is also a limit where dynamical instability for radial
perturbations sets in \cite{SWZ:81}.
 All the other configurations of a counterclockwise spiralling $\beta(E_r)$
curve are unstable \cite{Katz:78}.
 A detailed discussion of thermodynamic stability, fluctuations and
phase transitions in relativistic radiation with $\eta=0$ has been given in
Parentani et al. \cite{PKO:94}.
 Here we are interested to know how stability limits change when one takes
account of the mass in the cavity ($\eta\neq 0$) and the field energy outside.

 Speaking of field energy, we are aware that the subject is controversial and,
of course, we take sides.
 $E_r$ is surely not $M-M_s$ because $M-M_s>0$ even when there is no radiation.
 The difference must be in the field energy outside the cavity $E_f$
($r\geq R$).
 To calculate $E_f$, we use Lynden-Bell and Katz's method \cite{LbK:85}.
 We replace the cavity with the radiation by a new empty cavity of the same
radius $R$ and enough mass energy $E'_s (\neq E_s)$ to produce the same
gravitational field at $r \geq R$.
 Since the total mass energy is $M$ and the mass energy of the shell is $E'_s$,
the field energy $E_f=M-E'_s$ because the flat interior of the new cavity has
{\it no energy}.
 $E_f/R$ depends on $M/R$ or $\alpha$ only
\begin{equation}
  \frac{E_f}{R} = \frac 12 \left[1-\sqrt{1-\frac{2m}{R}} \;\right]^2
                = \frac 12 (1-z)^2.
\label{eq:defEf}
\end{equation}
 Thus, radiation energy or
\begin{equation}
    \varepsilon_r \equiv \frac{E_r}{R} = \alpha - \frac{E_s}{R} - \frac{E_f}{R}
\label{eq:defvarE1}
\end{equation}
with $E_s/R$ given in (\ref{eq:EsLEa}) and $y, z$ defined in (\ref{eq:defyz}),
one can write
\begin{equation}
  \varepsilon_r = z(1-y) = \frac 12 (2-\eta) (1-z) (z-z_1).
\label{eq:defvarE2}
\end{equation}
 $\varepsilon_r$ is zero at $z=z_1$ and $z=1$ and has one maximum
$\varepsilon_{rmax}$ at
\begin{equation}
    z' = \frac{1+z_1}{2} \;\;\;\;\; \mbox{where} \;\;\;\;\;
   \varepsilon_{rmax} = \frac{(1-\eta)^2}{2(2-\eta)}.
\label{eq:zM}
\end{equation}
 The $\beta(\varepsilon_r)$ curves are inward spirals for $\eta<\eta_0$ or
U-type like curves for $\eta>\eta_0$ similar to $\beta(\mu)$ shown in figure 3
and 4.
 Assuming radiation is thermodynamically stable at low temperature
($\beta\rightarrow\infty$) all configurations for which
$0<\varepsilon_r<\varepsilon_{rmax}$ and $\beta$ has its lowest value are
stable\cite{Katz:78}.
 The spiral of $\beta(\varepsilon_r)$ at higher values of $\beta$ represent
unstable configurations.
 The limits of stability in the $\beta(\mu)$ plane depend on the value of
$\eta$.

 We consider first $\eta>\eta_0$ for which $z_1\leq z \leq 1$ and the $y(z)$
line for $\eta>\eta_0$ in figure 2.
 Starting from $z=1$ ($\mu=0$, zero mass energy), we pump in slowly energy,
evolving through a succession of quasi-equilibrium configurations.
 This decreases both $z$ and $y$.
 As we know, $y$ reaches a minimum (or $\mu$ a maximum) at
$z=\sqrt{z_1}<1$.
 But $\sqrt{z_1}<\frac{1+z_1}{2}=z'$; the radiation energy reaches thus its
maximum for $\mu=\mu'<\tilde\mu<\mu_{max}$.
 At that point obtained from $y(z')=y'$,
\begin{equation}
  \mu' = \frac 12 (1-\eta)^2 - \frac 18 (1-\eta)^4
\label{eq:defmuprime}
\end{equation}
 and since $\eta>\eta_0$,
\begin{equation}
  \mu'<\frac 12 (1-\eta_0)^2 - \frac 18 (1-\eta_0)^4
    = \frac 14 - \frac {1}{32} = \frac {7}{32}
\label{eq:ineqmuprime}
\end{equation}
 Thus, in a linear series of increasing energy, the radiation becomes unstable
at $\mu=\mu'$ before reaching $\mu=\tilde\mu(\eta)<\mu_{max}$, the higher
the mass energy of the cavity, the smaller $\mu'$.
 The limit of stability in the $\beta(\mu)$ plane is shown in figure 4.

 Next consider $\eta<\eta_0$ with $z_3\leq z\leq 1$ for which there is a small
region in the $(\alpha,\mu)$ plane where the dominant energy condition holds
(see figure 5).
 When $0.235\leq \eta<\eta_0$, $z'$ is greater than $z_3$ and $\varepsilon_r$
reaches its maximum for a $\mu'<\mu_{max}$.
 For $0\leq\eta<0.235$, $z_3$ is greater than $z'$ and $\varepsilon_r$ reaches
its
maximum for $\mu'=\mu_{max}$.
 The limits of stability, where $\varepsilon_r$ is maximum, are shown in
figure 3a.

 Finally, in the case where $\eta<\eta_0$ but $z_1\leq z\leq z_2$ the maximum
of $\varepsilon_r$ is at $z=z_2$ where $\mu=\mu_{max}$ because $z_2$ is always
smaller than $z'$.
 The limit of stability is also shown in figure 3b.

\vspace{0.6cm}
\section{Concluding Remarks}

 It is clear that most cavities containing thermal radiation and having less
than $\simeq 30\%$ of the total mass energy are not realistic in the sense
that they have to sustain too high tensions or pressures.
 Those cavities, able to sustain such tensions or pressures contain, however,
in most cases, radiation that is thermodynamically unstable and would
collapse to form a black hole\cite{GP:78}.
 In particular, all cavities with negligible mass energy ($\eta=0$) that are
viable contain unfortunately unstable radiation as can be seen in figure 3b.
 Thus, the point raised by Sorkin et al.\cite{SWZ:81} about total entropy
becoming infinite in cavities with increasing radius or even reaching the
Bekenstein \cite{Bek:81} limit $S=2\pi RM$ does not apply to stable radiation
in \lq viable\rq \ cavities.

 One can have some fun by putting cavities with radiation on the other side of
the Einstein-Rosen bridge in a \lq perpetual\rq \ Schwarzschild spacetime.
 In this case, $\sigma$ is higher and $\Pi$ is always negative and instead of
(\ref{eq:defsig}) and (\ref{eq:defpi}) one has now
\begin{equation}
 4 \pi G R \sigma = \sqrt{1-2\mu} + \sqrt{1-2 \alpha}
\label{eq:defsigER}
\end{equation}
and
\begin{equation}
 4 \pi G R \Pi = -\frac 12
   \left [ \frac{1-\alpha}{\sqrt{1-2 \alpha}}
          +\frac{1-\mu+q(\mu)/3}{\sqrt{1-2 \mu}} \right].
\label{eq:defpiER}
\end{equation}
 Viable cavities are for $\Pi>-\sigma$.
 One can also put a small black hole in the center like Hawking \cite{Haw:76}
and look for the changes our considerations would make in a thermodynamic
stability analysis (Parentani et al.\cite{PKO:94}) if one wishes to be
realistic about probabilities of phase transition, i.e. probabilities for
black body radiation to collapse and form a black hole \cite{GP:78}.

\newpage

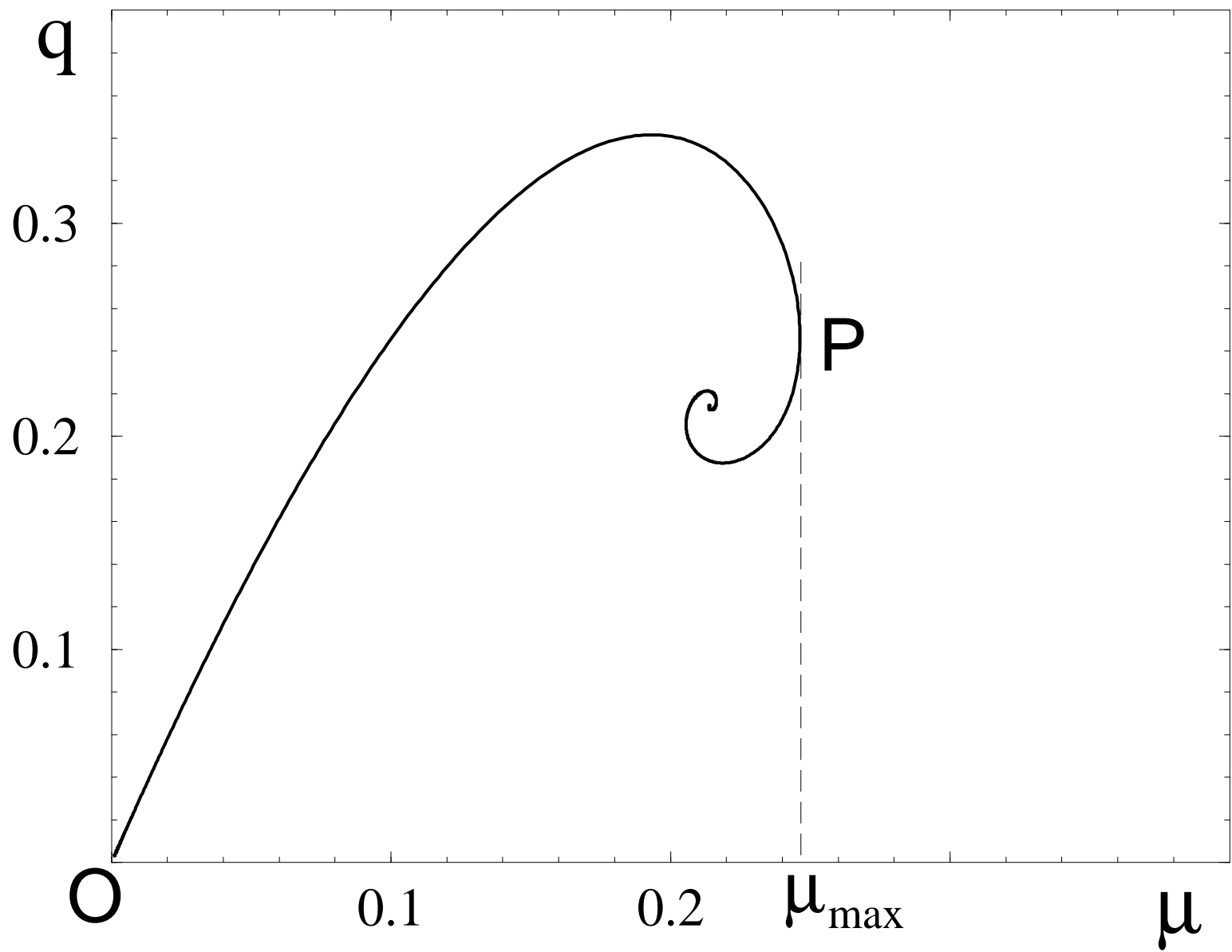
Fig.1: This represents the solution $q(\mu)$ of the TOV equation (12) and
$\mu_{max}\simeq 0.246$.\vspace{0.5cm}

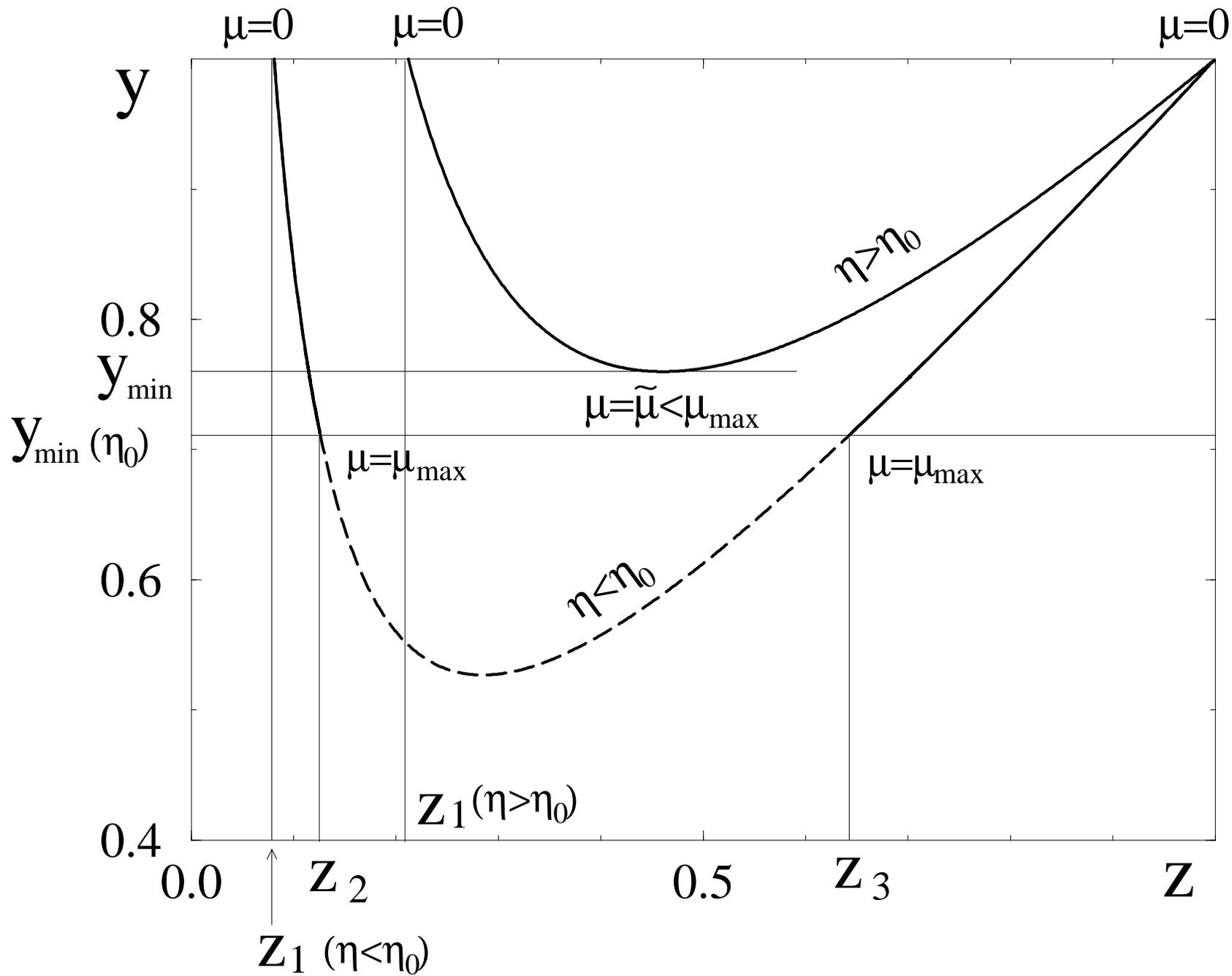
Fig.2: Two $y(z)$ lines, parametrized in $\mu$:
 one for $\eta=0.35>\eta_0\simeq 0.29$ whose minimum $y_{min}$ is at
$z=\sqrt{z_1(\eta)}$ for $\mu=\tilde\mu<\mu_{max}$ and
 another for $\eta=0.15<\eta_0\simeq 0.293$ whose minimum at
$z=\sqrt{z_1(\eta)}$ is below $y_{min}(\eta_0)$.\vspace{0.5cm}

Fig.3a: The dark region in this $B(\mu)$ plane corresponds to
$|\Pi|\leq\sigma$ when $z_1<z<z_2$ and holds only for
$0.25\leq\eta\leq\eta_0\simeq 0.29$.
 There are no viable cavities for $\eta<0.25$. The two linear series in
this figure are in solid lines where configurations are stable (section 7)
and dashed for unstable ones.\vspace{0.5cm}

Fig.3b: Dark regions in the plane correspond to $|\Pi|\leq\sigma$ when
$z_3\leq z\leq 1$ and holds for $0<\eta<\eta_0$.
 the two linear series in the figure are in solid lines where stable
(section 7) and dashed where unstable.
 There are no {\it stable} lines in the dark region for $\eta\leq 0.11$.
 The dot-dashed line represents $\eta=0$ configurations.
 As can be seen, such cavities either contain unstable radiation or are unable
to sustain the pressures.\vspace{0.5cm}

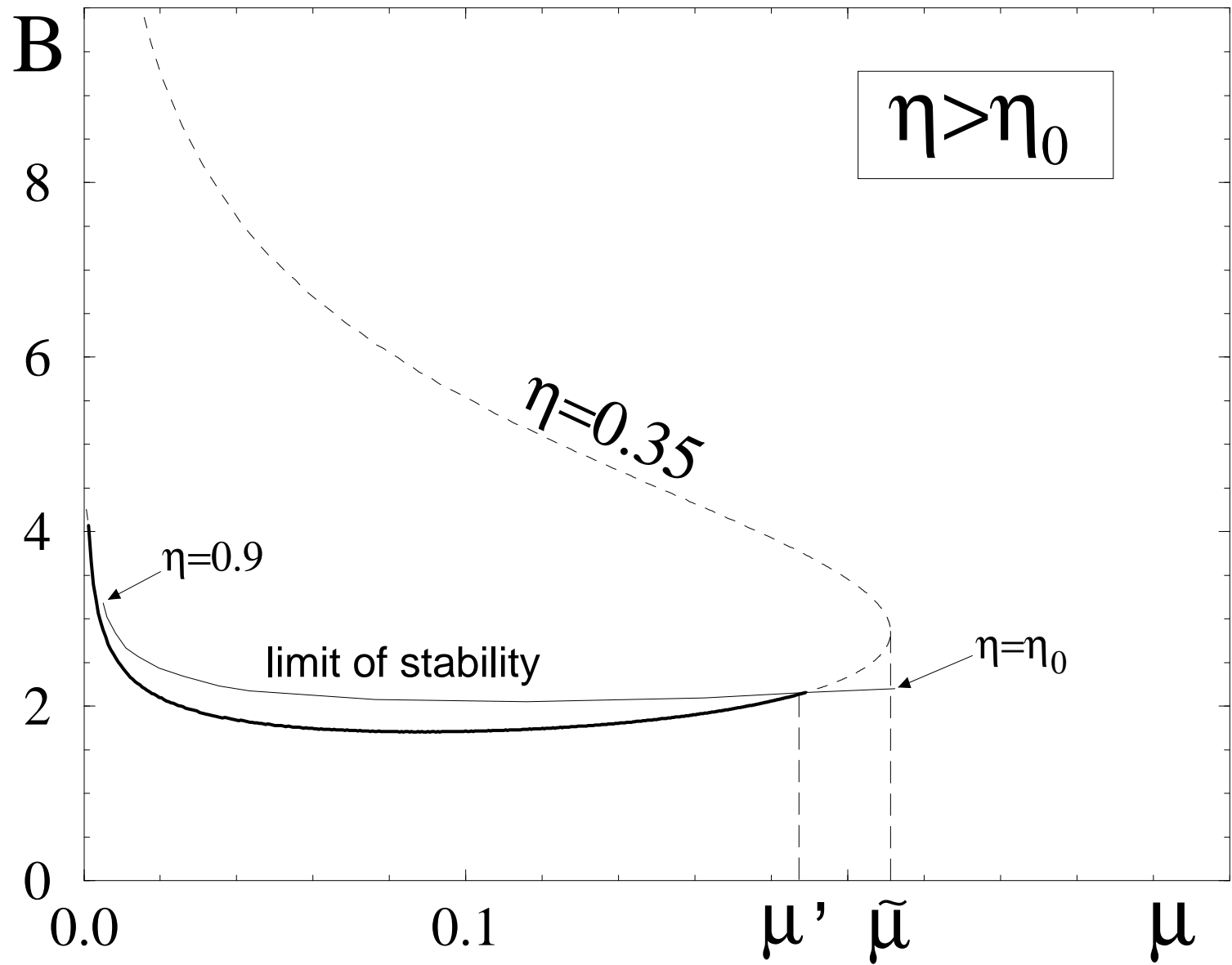
Fig.4: The linear series $B(\mu)$ for $\eta=0.35$ is
characteristic of all lines for $\eta>\eta_0$. The dashed part of the line
represents unstable configurations (section 7) the thick lines are for stable
ones.
 All configurations are unstable above the thin continuous line representing
the limit of stability. \vspace{0.5cm}

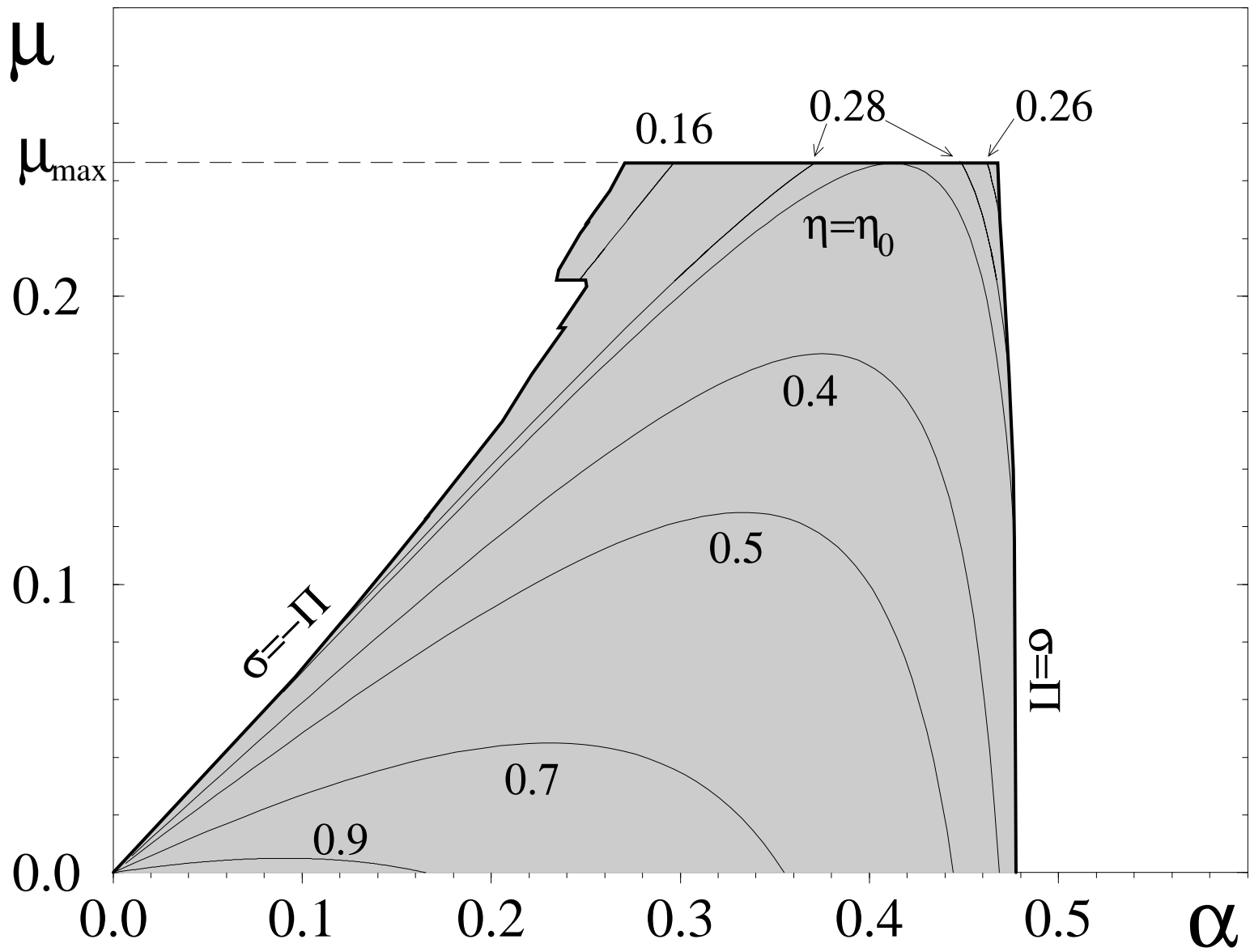
Fig.5: The dark region of the $(\alpha,\mu)$ plane are points associated with
equilibrium radiation in cavities satisfying the dominant energy conditions.

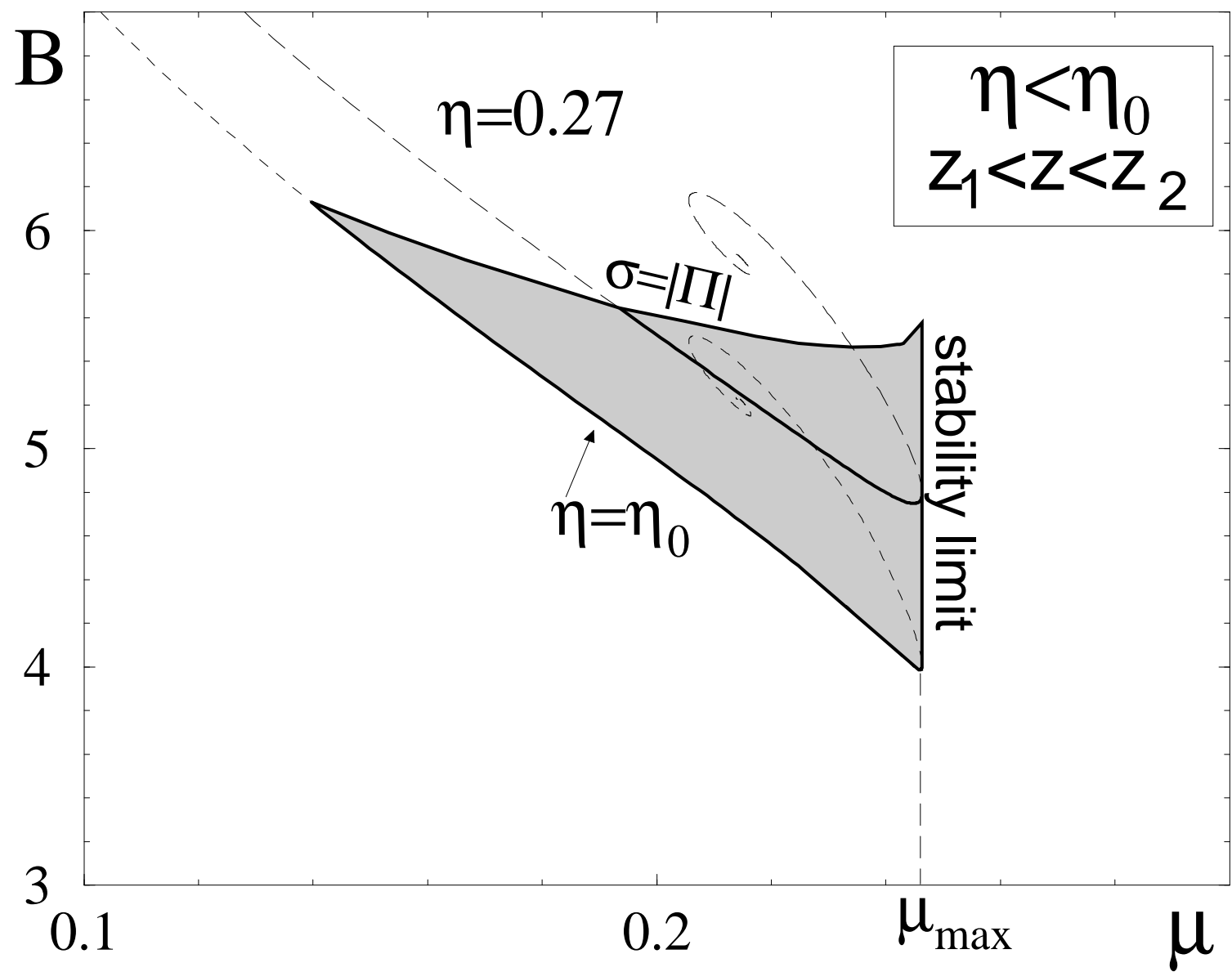
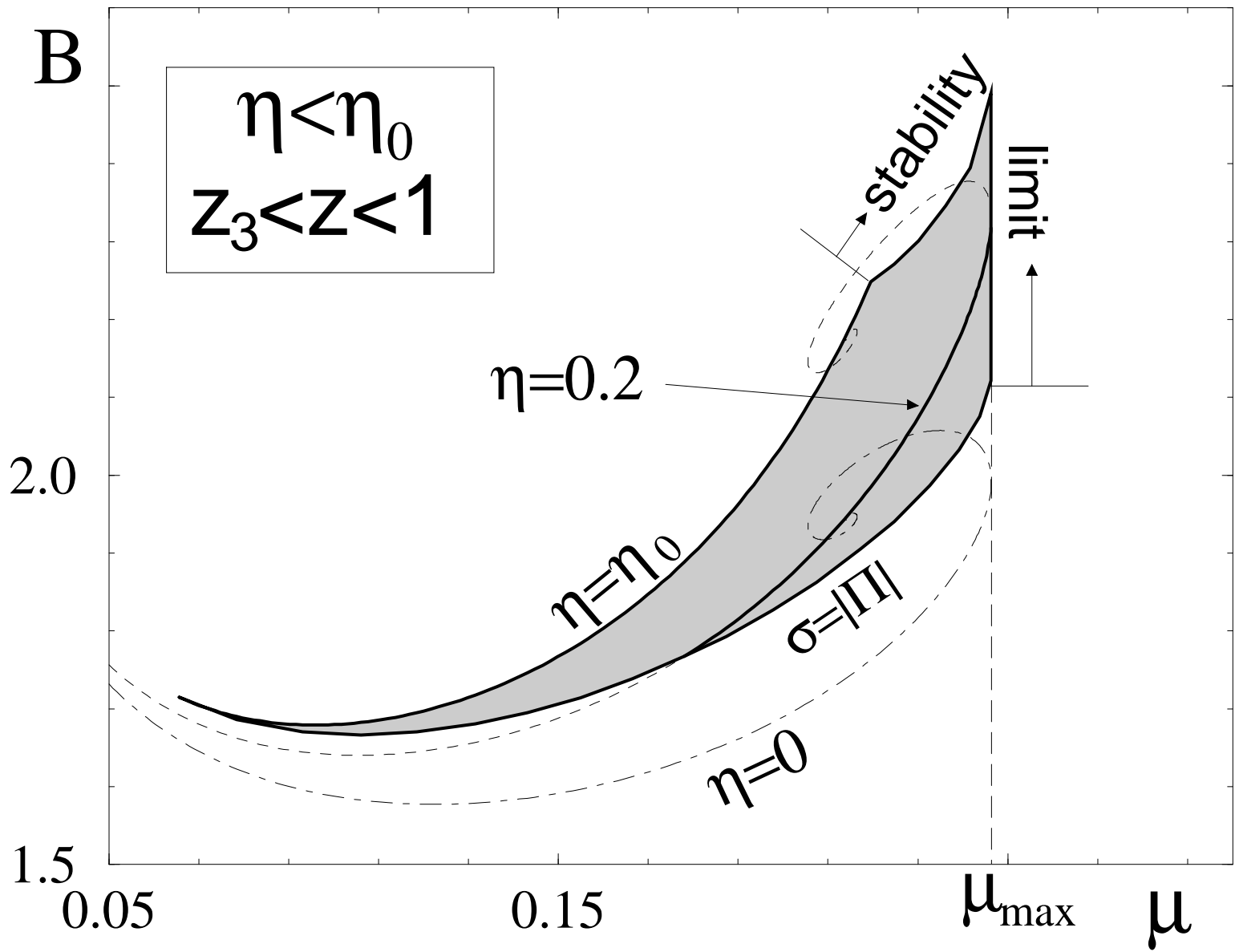
\end{document}